\documentstyle[aps,prd,epsfig,twocolumn,floats]{revtex}

\newcommand{\Tr}{{\mathrm{Tr}}}

\newcommand{\Asl}{\rlap{\,/}{{A}}}
\newcommand{\Dsl}{\rlap{\,/}{{D}}}
\newcommand{\pasl}{\rlap{/}{{\partial}}}

\newcommand{\RMT}{\mathrm{RMT}}
\newcommand{\disc}{\mathrm{disc}}
\newcommand{\chidisc}{\chi^{\disc}}
\newcommand{\eqref}[1]{(\ref{#1})}

\begin{document}            % the beginning
\draft
\wideabs{
\title{Small eigenvalues of the SU(3) Dirac operator\\ on the lattice
  and in Random Matrix Theory}
\author{M.~G\"ockeler, H.~Hehl, P.E.L.~Rakow, A.~Sch\"afer}
\address{Institut f\"ur Theoretische Physik,
  Universit\"at Regensburg, D-93040 Regensburg, Germany}
\author{T.~Wettig}
\address{Institut f\"ur Theoretische Physik,
  Technische Universit\"at M\"unchen, D-85747 Garching, Germany}
\date{November 13, 1998}
\maketitle
\begin{abstract}
  We have calculated complete spectra of the staggered Dirac operator
  on the lattice in quenched SU(3) gauge theory for $\beta = 5.4$ and
  various lattice sizes.  The microscopic spectral density, the
  distribution of the smallest eigenvalue, and the two-point spectral
  correlation function are analyzed.  We find the expected agreement
  of the lattice data with universal predictions of the chiral unitary
  ensemble of random matrix theory up to a certain energy scale, the
  Thouless energy.  The deviations from the universal predictions are
  determined using the disconnected scalar susceptibility.  We find
  that the Thouless energy scales with the lattice size as expected
  from theoretical arguments making use of the
  Gell-Mann--Oakes--Renner relation.
\end{abstract}
}

\narrowtext

The low-lying eigenvalues of the Dirac operator are of great
importance for the understanding of spontaneous chiral symmetry
breaking in an infinite volume \cite{Bank80}.  On the lattice,
however, one is always working at finite volume. Therefore, it is
important to know how the thermodynamic limit is approached.  It was
shown by Leutwyler and Smilga \cite{Leut92} that in the domain
\begin{equation}
\label{domain}
  1/\Lambda\ll L\ll1/m_\pi\:,   
\end{equation}
where $\Lambda$ is a typical hadronic scale, $L$ is the linear extent
of the Euclidean box, and $m_\pi$ is the pion mass, the low-energy
behavior of QCD can be described by a simple effective partition
function whose existence imposes certain constraints on the
eigenvalues of the Dirac operator.  The spectrum of the Dirac operator
in the domain \eqref{domain} has been successfully predicted by chiral
random matrix theory (RMT) \cite{Shur93,Verb93}.  The only ingredients
of the calculation are the global symmetries of the theory and the
assumption that chiral symmetry is spontaneously broken. It has
recently been shown that there is an overlap between the domain of
validity of chiral RMT and of chiral perturbation theory, and that in
this overlap region the two approaches yield the same results
\cite{Osb98}.

The results so obtained provide analytical information on the way in
which the thermodynamic limit is approached.  They are universal in
the sense that they do not depend on the precise values of the
parameters of the theory, i.e., of the simulation parameters on the
lattice.  However, the domain of validity of the universal results
does depend on the parameters.  The energy scale $\lambda_{\RMT}$ up
to which RMT applies, i.e., the Thouless energy, follows from the
upper bound on $L$ in relation~\eqref{domain} and the
Gell-Mann--Oakes--Renner relation, $m_\pi^2f_\pi^2=2m\Sigma$, where
$f_\pi=F_\pi/\sqrt{2}$ is the pion decay constant, $\Sigma$ is the
absolute value of the chiral condensate $\langle\bar\psi\psi\rangle$,
and $m$ is a valence quark mass. It is thus determined by
\cite{Jani98,Osbo98a,Ster98}
\begin{equation}
  \label{thouless}
  \lambda_{\RMT}/\Delta\propto f_\pi^2L^2\:,
\end{equation}
where $\Delta=1/\rho(0)=\pi/(V\Sigma)$ is the level spacing at zero.
Here, $V=L^4$ denotes the four-volume, and
$\rho(\lambda)=\langle\sum_n\delta(\lambda-\lambda_n)\rangle_A$ is the
spectral density of the Dirac operator averaged over gauge field
configurations $A$.  The relation between $\rho(0)$ and $\Sigma$ is
given by the Banks--Casher formula, $\pi\rho(0)=V\Sigma$
\cite{Bank80}.

The aim of this paper is (i) to test the universal predictions of
chiral RMT for the distribution and correlations of the low-lying
Dirac eigenvalues and (ii) to check the prediction of
Eq.~\eqref{thouless} for the Thouless energy, using lattice data
computed in quenched SU(3) gauge theory with the staggered Dirac
operator.  Point (i) has previously been considered in quenched SU(2)
\cite{Berb98a,Ma98}, in SU(2) with dynamical fermions \cite{Berb98b},
in quenched SU(3) in three dimensions \cite{Damg98a}, in U(1) in two
dimensions \cite{Farc98}, and, very recently, in quenched SU(3) in
four dimensions \cite{Damg98b}.  Point (ii) has previously been tested
in quenched SU(2) \cite{Berb98c}.  All these investigations were done
with the staggered Dirac operator except for Ref.~\cite{Farc98} in
which the fixed point Dirac operator with respect to a renormalization
group transformation was used.  Since real QCD has three colors, SU(3)
in four dimensions is clearly the most important case.

The Euclidean Dirac operator in the continuum is given by $i\Dsl =
i\pasl+gt^a\Asl^a$, where the $t^a$ are the generators of the gauge
group.  The operator $i\Dsl$ is hermitean with real eigenvalues. It
anticommutes with $\gamma_5$ and, therefore, all nonzero eigenvalues
come in pairs $\pm\lambda_n$ with eigenvectors $\psi_n,
\gamma_5\psi_n$.  There can also be zero modes which are either
left-handed or right-handed.  The topological charge of a given gauge
field configuration is equal to the difference in the number of
left-handed and right-handed zero modes.  On the lattice, the
staggered Dirac operator reads
\begin{eqnarray}
(i\Dsl)_{x,y} = \frac i2 \sum_\mu\,\bigl[&&\eta_\mu(x)U_\mu(x)\cdot
  \delta_{x+\mu,y}\nonumber \\
  &&-\eta_\mu(y)U^\dag_\mu(y)\cdot\delta_{x-\mu,y}\bigr]\:,
\end{eqnarray}
where $U$ and $\eta$ denote the link variables and the staggered
phases, respectively.

The claim is that the distribution and the correlations of the small
eigenvalues of $i\Dsl$ are described by universal functions which can
be computed, e.g., in chiral RMT \cite{Shur93,Verb93}.  The
distribution of the low-lying eigenvalues is encoded in the spectral
one-point function near zero virtuality, the so-called microscopic
spectral density defined by \cite{Shur93}
\begin{equation}
  \label{rhos}
  \rho_s(z) = \lim_{V\to\infty}\frac{1}{V\Sigma}\,
  \rho\left(\frac{z}{V\Sigma}\right)\:.
\end{equation}
Similarly, one considers the microscopic limit of the two-point
cluster function,
\begin{equation}
  \label{tau2}
  \tau_2(z_1,z_2)=\lim_{V\to\infty}\frac{1}{(V\Sigma)^2}\,T_2\left(
  \frac{z_1}{V\Sigma},\frac{z_2}{V\Sigma}\right)
\end{equation}
with
\begin{equation}
  T_2(\lambda_1,\lambda_2) = \rho(\lambda_1)\rho(\lambda_2) -
  R_2(\lambda_1,\lambda_2)\:,
\end{equation}
where $R_2(\lambda_1,\lambda_2)$ is the two-point spectral correlation
function, i.e., the probability density that one eigenvalue is at
$\lambda_1$ and another at $\lambda_2$, all other eigenvalues being
unobserved.  

For SU(3) with the staggered Dirac operator, the relevant symmetry
class in the framework of chiral RMT is the chiral unitary ensemble
\cite{Verb94a}.  In the following, we briefly summarize analytical
results for this ensemble which are of relevance for the present work.
The microscopic spectral density is given by \cite{Verb93}
\begin{equation}
  \label{rhosrmt}
  \rho_s(z)=\frac z2\left[J_\mu^2(z)-J_{\mu+1}(z)J_{\mu-1}(z)\right]
\end{equation}
with the Bessel function $J$ and $\mu=N_f+|\nu|$, where $N_f$ and
$\nu$ denote the number of massless flavors and the topological
charge, respectively.  The distribution of the smallest eigenvalue for
$N_f=\nu=0$ reads \cite{Forr93}
\begin{equation}
  \label{plmin}
  P(\lambda_{\min}) =
  \frac{\lambda_{\min}}{2}\:e^{-\lambda_{\min}^2/4}\:.
\end{equation}
The two-point cluster function in the microscopic limit is given by
\cite{Verb93} 
\begin{eqnarray}
  \label{tau2rmt}
  \lefteqn{\tau_2(z_1,z_2)} \nonumber\\
  &=&z_1z_2\left[\frac{z_1J_{\mu+1}(z_1)J_\mu(z_2) - 
      z_2J_\mu(z_1)J_{\mu+1}(z_2)}{z_1^2 - z_2^2}\right]^2\:.
\end{eqnarray}
The quantities in Eqs.~\eqref{rhosrmt} through \eqref{tau2rmt} do not
contain any free parameters.  For a comparison with lattice data, the
energy scale is determined by the parameter $V\Sigma$ which is
obtained from the data by extracting $\rho(0)$ and applying the
Banks--Casher relation, $\pi\rho(0)=V\Sigma$.  Thus, the comparison
between lattice data and the predictions of Eqs.~\eqref{rhosrmt}
through \eqref{tau2rmt} is parameter-free.  (Strictly speaking, on
finite lattices a spontanous breaking of chiral symmetry cannot occur
and $\rho(0)$ is zero.  The latter quantity must, therefore, be
determined by extrapolating $\rho(\lambda)$ to $\lambda=0$.  In
practice, this extrapolation presents no difficulties.)

We now turn to the details of our numerical simulations.  They were
done in quenched SU(3) gauge theory with $\beta=6/g^2=5.4$ on lattices
of size $V=L^4$ with $L=4$, 6, 8, 10.  The boundary conditions are
periodic for the gauge fields and periodic in space and anti-periodic
in Euclidean time for the Dirac operator.  The gauge field
configurations were generated using a combined Metropolis and
overrelaxation algorithm on the link variables.  Two consecutive
configurations are separated by at least 30 runs of one Metropolis
sweep with three hits and 20 overrelaxation sweeps using Creutz's
method \cite{Creutz87}. The complete spectrum of the staggered Dirac
operator was then calculated using the Cullum--Willoughby version of
the Lanczos algorithm for the matrix of $-\Dsl^2$.  This operator
couples only even to even and odd to odd lattice sites.  Both blocks
have the same eigenvalues. Hence it is sufficient to consider only
even lattice sites.  The eigenvalues of $-\Dsl^2$ were checked against
the identity $\Tr(-\Dsl^2) = 3V$ which was fulfilled with relative
accuracy $10^{-9}$.  The total number of diagonalized configurations
and the extrapolated values of $\pi\rho(0)=V\Sigma$ are shown in
Table~\ref{table}.
\begin{table}
  \caption{Simulation parameters and extrapolated value of $V\Sigma$
    for $\beta = 5.4$.}
  \vspace*{2mm}
  \begin{tabular}{rrr@{$\,\pm\,$}r}
    $L$ & conf. & \multicolumn{2}{c}{$V\Sigma$}
    \\[0.5mm]\tableline\\[-3mm]
    4 & 35337 & 225 & 7\\
    6 & 11748 & 1207 & 28\\
    8 & 2635 & 3918 & 58\\
    10 & 1059 & 9429 & 155\\
  \end{tabular}
  \label{table}
\end{table}

\begin{figure}
  \centerline{\epsfig{file=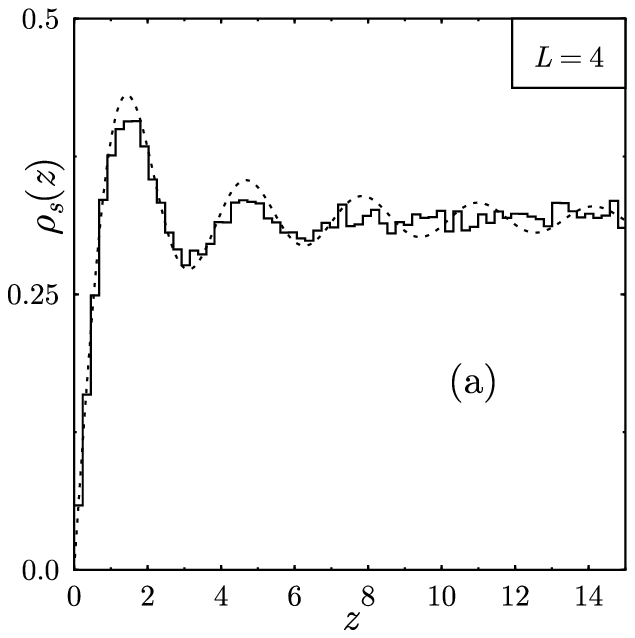, width = 47.5mm,
      height=52mm}\hspace*{-5mm}
    \epsfig{file=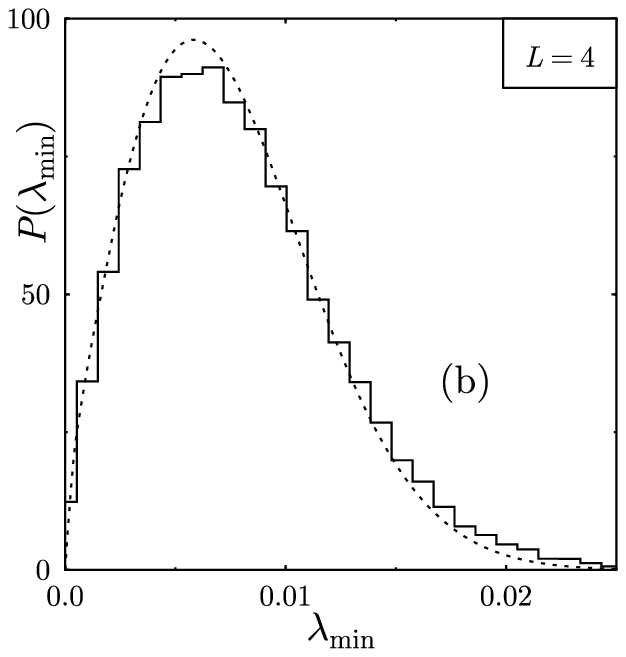, width = 47.5mm,
      height=52mm}}%\\\vspace*{-2mm}
    \centerline{\epsfig{file=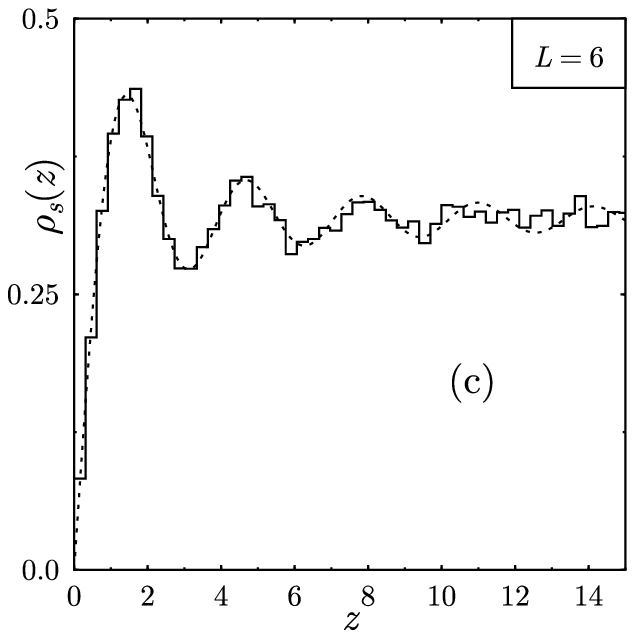, width = 47.5mm,
      height=52mm}\hspace*{-5mm}
    \epsfig{file=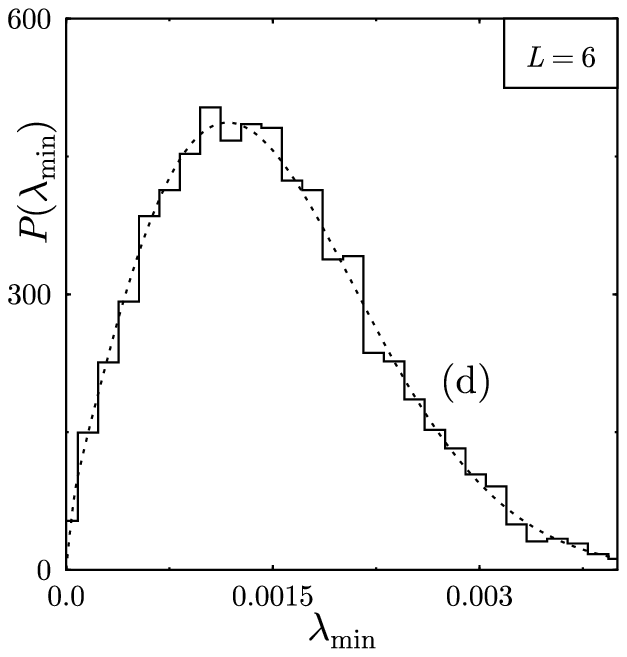, width = 47.5mm,
      height=52mm}}%\\\vspace*{-2mm}
    \centerline{\epsfig{file=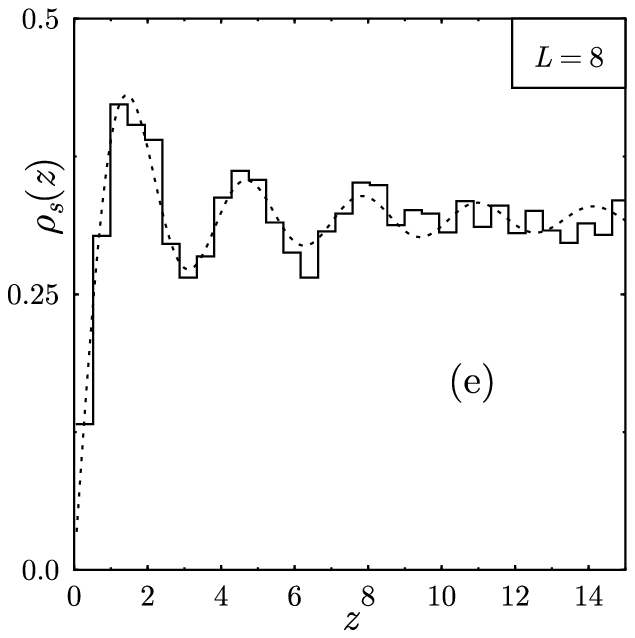, width = 47.5mm,
      height=52mm}\hspace*{-5mm}
    \epsfig{file=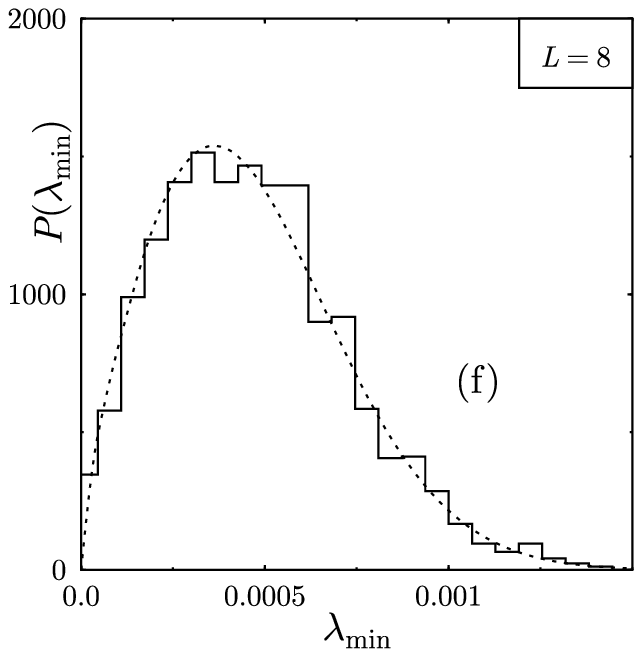, width = 47.5mm,
      height=52mm}}%\\\vspace*{-2mm}
    \centerline{\epsfig{file=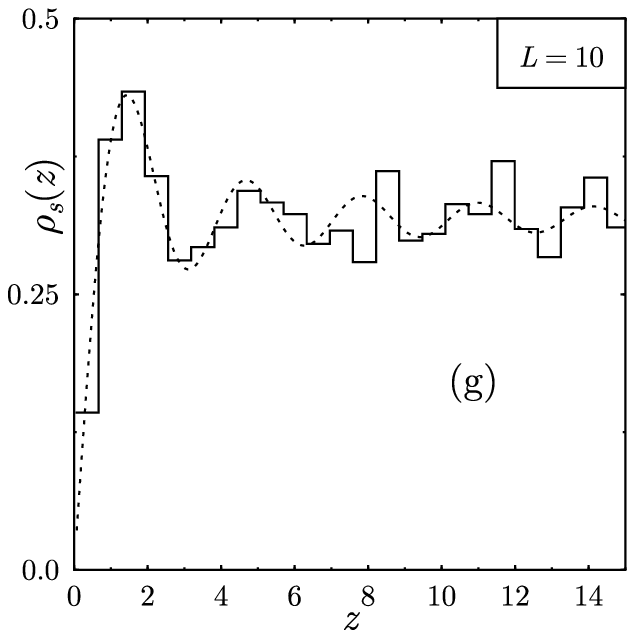, width = 47.5mm,
      height=52mm} \hspace*{-5mm}
    \epsfig{file=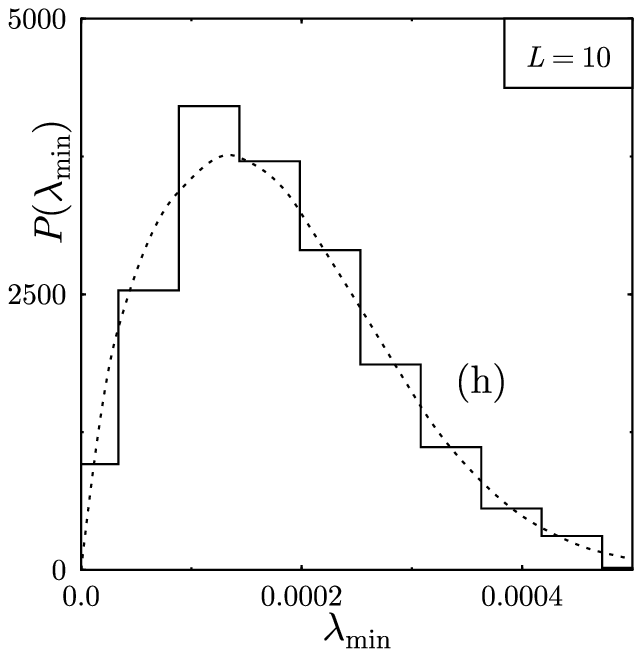, width = 47.5mm,
      height=52mm}}
  \vspace*{2mm}
  \caption{Microscopic spectral density (left) and distribution of
    the smallest eigenvalue (right) of the staggered Dirac operator in
    quenched SU(3) for $\beta=5.4$ and $V=L^4$ with $L=4$, 6, 8, 10.
    The histograms represent the lattice data, the dashed curves are
    the predictions of Eqs.~(\protect\ref{rhosrmt}) (with $\mu=0$) and
    (\protect\ref{plmin}).} 
  \vspace*{2mm}
  \label{rhoandP}
\end{figure}

The lattice data for $\rho_s(z)$ and $P(\lambda_{\min})$ are compared
with the predictions of Eqs.~\eqref{rhosrmt} and \eqref{plmin} in
Fig.~\ref{rhoandP}.  
In Eq.~\eqref{rhosrmt}, we have used $\mu=0$.  Clearly, $N_f=0$ since
we consider the quenched approximation.  The fact that $\nu=0$ is less
obvious.  The prediction of Eq.~\eqref{rhosrmt} is restricted to
sectors with definite topological charge.  Therefore, one should
compute the topological charge of each gauge field configuration and
compare the lattice data in each topological sector with the
prediction of Eq.~\eqref{rhosrmt}.  However, Eq.~\eqref{rhosrmt}
assumes that for $\nu\ne0$ the Dirac operator has exact zero modes.
This is not the case for staggered fermions on the lattice where at
finite lattice spacing $a$ the would-be zero modes are shifted by an
amount proportional to $a^2$ \cite{Vink88}.  For the value of $\beta$
we used, $a$ is still relatively large so that no zero modes are
present.  This explains why the lattice data are consistent with
Eq.~\eqref{rhosrmt} for $\nu=0$, as seen in Fig.~\ref{rhoandP}. Very
similar results for different $\beta$ were very recently presented in
\cite{Damg98b}.

\begin{figure}
  \centerline{\epsfig{file=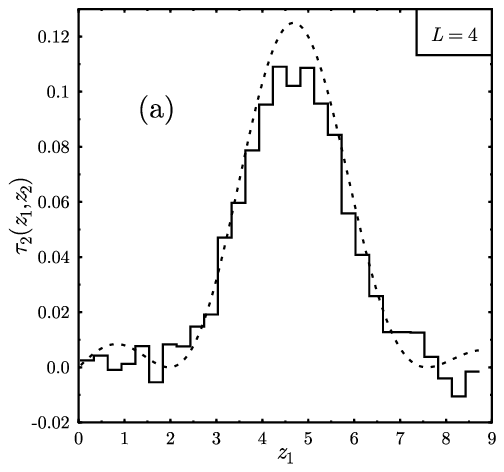, width=51mm}\hspace*{-8mm}
    \epsfig{file=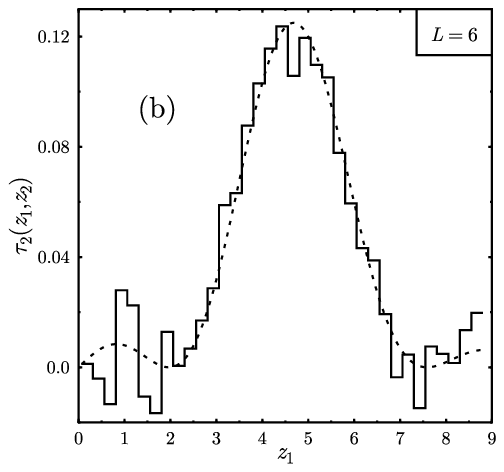, width=51mm}}
  \caption{Microscopic two-point cluster function $\tau_2(z_1,z_2)$
    with $z_2$ at the second maximum of $\rho_s(z)$ for $L=4$ and
    $L=6$, respectively.  The histograms represent the lattice data,
    the dashed curves are the prediction of
    Eq.~(\protect\ref{tau2rmt}) for $\mu=0$.}
  \label{tau2fig}
\end{figure}

The agreement between the lattice data and the universal predictions
is quite satisfactory, also for the two-point cluster function in the
microscopic limit which we have plotted in Fig.~\ref{tau2fig} along
with the prediction of Eq.~\eqref{tau2rmt} for $\mu=0$.  
A fixed value of $z_2=4.68$ (corresponding to the location of the
second maximum of $\rho_s(z)$) was chosen.  

The quantity $\tau_2(z_1,z_2)$ is interesting since it enters in the
calculation of the disconnected scalar susceptility which, in turn,
can be used to determine the Thouless energy, i.e., the scale
$\lambda_{\RMT}$ above which the lattice data deviate from the
universal predictions of Eqs.~\eqref{rhosrmt} through \eqref{tau2rmt}.
In terms of the Dirac eigenvalues, this quantity is defined as
\cite{Verb96}
\begin{eqnarray}
  \label{chidisclatt}
  \chi^{\disc} &= &\frac{1}{V}\left\langle\sum_{k,\ell=1}^N
    \frac{1}{(i\lambda_k+m)(i\lambda_\ell+m)}\right\rangle\nonumber\\
  &&\mbox{}-\frac{1}{V}\left\langle\sum_{k=1}^N
    \frac{1}{i\lambda_k+m}\right\rangle^2\:,
\end{eqnarray}
where $V$ is the number of lattice sites, $N$ the number of
eigenvalues and $m$ a valence quark mass, respectively.  The average
is over independent gauge field configurations.
Eq.~\eqref{chidisclatt} can be rewritten in terms of integrals
involving the spectral one- and two-point functions of the Dirac
operator.  Rescaling $\chi^{\disc}$ by $1/(V\Sigma^2)$ and changing
from $m$ to $u=mV\Sigma$, we have
\begin{eqnarray}
  \label{chirmt}
  \chidisc(u) &=& 4u^2\int_0^\infty dx\:\frac{\rho_s(x)}{(x^2+u^2)^2}
  \nonumber\\
  &&\mbox{}-4u^2\int_0^\infty\int_0^\infty dxdy \:
  \frac{\tau_2(x,y)}{(x^2+u^2)(y^2+u^2)} \nonumber\\
%  &=&2I_{\mu+1}(u)K_{\mu-1}(u)-\frac{4}{u^2}F_\mu(u) \:,
   &=&-u^2[K_\mu^2(u)-K_{\mu+1}(u)K_{\mu-1}(u)] \nonumber\\
   &&\quad\times[I_\mu^2(u)-I_{\mu+1}(u)I_{\mu-1}(u)]\:,
\end{eqnarray}
where in going from the first to the second line we have inserted the
RMT results for $\rho_s$ and $\tau_2$. %and
%\begin{equation}
%  F_\mu(u) = \int_0^u
%  dt\:t^3K_\mu^2(t)\bigl[I_\mu^2(t)-I_{\mu+1}(t)I_{\mu-1}(t)\bigr] \:.
%\end{equation}
The functions $I$ and $K$ are modified Bessel functions.
In the case of $\mu=0$, Eq.~\eqref{chirmt} simplifies to
\begin{equation}
  \chidisc(u) =
  u^2\bigl[K_1^2(u)-K_0^2(u)\bigr]\bigl[I_0^2(u)-I_1^2(u)\bigr] \:.
\end{equation}

\begin{figure}
  \centerline{\epsfig{file=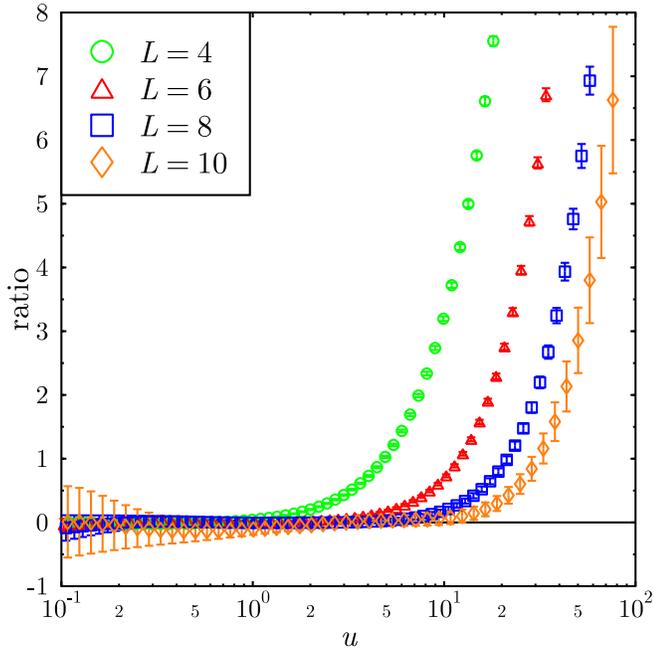,width=11cm}}
  \caption{The ratio defined in Eq.~(\protect\ref{ratio}) for
    $\beta=5.4$ and various lattice sizes.}
  \label{ratiofig}
\end{figure}

In order to compare the lattice results for $\chidisc$ obtained from
Eq.~\eqref{chidisclatt} with the RMT prediction of Eq.~\eqref{chirmt} we
introduce the variable \cite{Berb98c}
\begin{equation}
  \label{ratio}
  \mathrm{ratio} = \frac{\chidisc_{\mathrm{lattice}}-\chidisc_{\RMT}}
  {\chidisc_{\RMT}}
\end{equation}
which is plotted in Fig.~\ref{ratiofig}.
This ratio should be close to zero in the domain of validity of the
RMT predictions and deviate from zero at some value of $u=u_{\RMT}$
which corresponds to the Thouless energy.  (The deviations of the
ratio from zero for very small values of $u$ are artefacts of the
finite lattice size and of finite statistics.  This point was
discussed in Ref.~\cite{Berb98c}.)

\begin{figure}[btp]
  \centerline{\epsfig{file=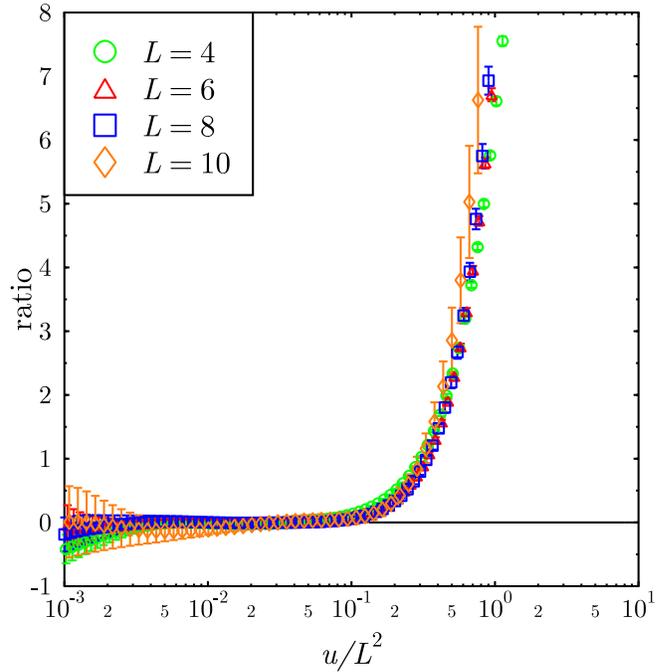,width=11cm}}
  \caption{The ratio of Eq.~(\protect\ref{ratio}) plotted versus
    $u/L^2$.} 
  \label{scalratio}
\end{figure}

The prediction of Eq.~\eqref{thouless} is that $\lambda_{\RMT}/\Delta$
should scale with $L^2$.  If we express $\lambda_{\RMT}$ in terms of
$u_{\RMT} = \lambda_{\RMT}V\Sigma = \pi\lambda_{\RMT}/\Delta$,
$u_{\RMT}$ should also scale with $L^2$.  To check this predicted
scaling behavior we have plotted the ratio of Eq.~\eqref{ratio} as a
function of $u/L^2$ in Fig.~\ref{scalratio}.
We observe that all data fall on the same curve, confirming the
prediction of Eq.~\eqref{thouless} with regard to the scaling with
$L^2$.  Since we have only considered one value of $\beta$, we cannot
check the scaling with $f_\pi^2$.  {}From Fig.~\ref{scalratio} we can
read off $\lambda_{\RMT}/\Delta\approx0.04L^2$ in lattice units.

In summary, we have shown that the distribution and the correlations
of the low-lying eigenvalues of the staggered Dirac operator in
quenched SU(3) are described by universal functions up to a certain
energy scale, the Thouless energy.  The latter quantity was determined
using the disconnected scalar susceptibility, and the predicted
scaling with $L^2$ was confirmed.  It would be of great interest to
extend the present study to dynamical fermions for which analytical
results are also available \cite{Damg97}.

{\bf Acknowledgments}.  We thank S. Meyer and H.A. Weidenm\"uller for
helpful discussions.  This work was supported in part by DFG grants
Scha-458/5-2 and We-655/15-1.


\begin{thebibliography}{99}
\bibitem{Bank80} T. Banks and A. Casher, Nucl. Phys. {\bf B169},
  103 (1980).
\bibitem{Leut92} H. Leutwyler and A.V. Smilga, Phys. Rev. D {\bf 46},
  5607 (1992).
\bibitem{Shur93} E.V. Shuryak and J.J.M. Verbaarschot, Nucl. Phys.
  {\bf A560}, 306 (1993).
\bibitem{Verb93} J.J.M. Verbaarschot and I. Zahed, Phys. Rev. Lett.
  {\bf 70}, 3852 (1993).
\bibitem{Osb98} J.C. Osborn, D. Toublan, and J.J.M. Verbaarschot,
  hep-th/9806110.
\bibitem{Jani98} R.A. Janik, M.A. Nowak, G. Papp, and I. Zahed,
  Phys. Rev. Lett. {\bf 81}, 264 (1998). 
\bibitem{Osbo98a} J.C. Osborn and J.J.M. Verbaarschot, Phys. Rev.
  Lett. {\bf 81}, 268 (1998); Nucl. Phys. {\bf B525}, 738 (1998).
\bibitem{Ster98} J. Stern, hep-ph/9801282.
\bibitem{Berb98a} M.E. Berbenni-Bitsch, S. Meyer, A. Sch\"afer, J.J.M.
  Verbaarschot, and T. Wettig, Phys. Rev. Lett. {\bf 80}, 1146 (1998).
\bibitem{Ma98} J.-Z. Ma, T. Guhr, and T. Wettig, Euro. Phys. J. A
  {\bf 2}, 87 (1998). 
\bibitem{Berb98b} M.E. Berbenni-Bitsch, S. Meyer, and T. Wettig,
  Phys. Rev. D {\bf 58}, 071502 (1998).
\bibitem{Damg98a} P.H. Damgaard, U.M. Heller, A. Krasnitz, and
  T. Madsen, hep-lat/9803012.
\bibitem{Farc98} F. Farchioni, I. Hip, C.B. Lang, and M. Wohlgenannt,
  hep-lat/9809049. 
\bibitem{Damg98b} P.H. Damgaard, U.M. Heller, and A. Krasnitz,
  hep-lat/9810060.
\bibitem{Berb98c} M.E. Berbenni-Bitsch, M. G\"ockeler, T. Guhr,
  A.D. Jackson, J.-Z. Ma, S. Meyer, A. Sch\"afer, H.A. Weidenm\"uller,
  T. Wettig, and T. Wilke, Phys. Lett. B {\bf 438}, 14 (1998).
\bibitem{Verb94a} J.J.M. Verbaarschot, Phys. Rev. Lett. {\bf 72},
  2531 (1994).
\bibitem{Forr93} P.J. Forrester, Nucl. Phys. {\bf B402}, 709 (1993).
\bibitem{Creutz87} M. Creutz, Phys. Rev. D {\bf 36}, 515 (1987).
\bibitem{Vink88} J.C. Vink, Phys. Lett. B {\bf 210}, 211 (1988).
\bibitem{Verb96} J.J.M. Verbaarschot, Phys. Lett. B {\bf 368}, 137
  (1996).
\bibitem{Damg97} P.H. Damgaard and S.M. Nishigaki, Nucl. Phys.
  {\bf B518}, 495 (1998);  T. Wilke, T. Guhr, and T. Wettig, Phys.
  Rev. D {\bf 57}, 6486 (1998);  G. Akemann and P.H. Damgaard, Phys.
  Lett. B {\bf 432}, 390 (1998);  B. Seif, T. Wettig, and T. Guhr,
  hep-th/9811044.
\end{thebibliography}
\end{document}